\documentstyle[a4,12pt]{article}
 
\newcommand{\newsection}[1]{
   \vspace{5mm}
   \pagebreak[3]
   \refstepcounter{section}
   \setcounter{equation}{0}
   \setcounter{subsection}{0}
   \setcounter{footnote}{0}   
\addcontentsline{toc}{section}{\protect\numberline{\arabic{section}}{#1}}
   \begin{center}
   {\bf {\large \thesection. #1}}
   \nopagebreak
   \end{center}
   \nopagebreak
   }
\newcommand{\newsubsection}[1]{
   \vspace{1cm}
   \pagebreak[3]
   \addtocounter{subsection}{1}
   \addcontentsline{toc}{subsection}{\protect
   \numberline{\arabic{section}.\arabic{subsection}}{#1}}
   \noindent{ \bf \thesubsection. #1}
   \nopagebreak
   \vspace{2mm}
   \nopagebreak}
\newcommand{\acknowledgements}{
   \vspace{15mm}
   \pagebreak[3]
   {\bf {\large Acknowledgements}}
   \nopagebreak
   \medskip
   \nopagebreak}
 
\newcommand{\ol}[1]{\overline{#1}}
\newcommand{\la}{\lambda}
\newcommand{\labar}{{\overline{\lambda}}}
\newcommand{\del}{\delta}
\newcommand{\ep}{\epsilon}
\newcommand{\epbar}{\overline{\epsilon}}

\newcommand{\Ga}[1]{\Gamma_{#1}}
\newcommand{\pd}[1]{\partial_{#1}}
\newcommand{\nn}{\nonumber}

\newcommand{\bvec}{\left(\begin{array}{c}}
\newcommand{\evec}{\end{array}\right)}
\newcommand{\bmat}{\left(\begin{array}{cc}}
\newcommand{\emat}{\end{array}\right)}
\newcommand{\fr}{\frac}
\newcommand{\be}{\begin{equation}}
\newcommand{\ee}{\end{equation}}
\newcommand{\ba}{\begin{eqnarray}}
\newcommand{\ea}{\end{eqnarray}}
\newcommand{\CL}{{\cal L}}
\newcommand{\Tr}{\mbox{Tr}\,}   

\def\NPB#1#2#3{{\it Nucl.\ Phys.}\/ {\bf B#1} (19#2) #3}
\def\PLB#1#2#3{{\it Phys.\ Lett.}\/ {\bf B#1} (19#2) #3}
\def\PRD#1#2#3{{\it Phys.\ Rev.}\/ {\bf D#1} (19#2) #3}
\def\PRL#1#2#3{{\it Phys.\ Rev.\ Lett.}\/ {\bf #1} (19#2) #3}

\def\JHEP#1#2#3{{\it JHEP}\/ {\bf #1} (19#2) #3}
\def\hep#1{\texttt{hep-th/#1}}

\begin{document}
 
\addtolength{\baselineskip}{.5mm} 
\renewcommand{\theequation}{\thesection.\arabic{equation}}
\renewcommand{\thefootnote}{\fnsymbol{footnote}}

\begin{flushright}
May 1999\\
SPIN-99/14 \\
hep-th/9905068\\
\end{flushright}
\vspace{2cm}
\thispagestyle{empty}
\begin{center}
{\large{U-Duality in Supersymmetric Born-Infeld Theory}}\\[18mm]
{\bf Gysbert Zwart\footnote{e-mail address: zwart@phys.uu.nl} }\\[7.5mm]
{\it Spinoza Institute\\ University of Utrecht\\P.O. Box 80 195\\
3508 TD Utrecht\\The Netherlands\\
\vspace{.5cm}
and\\
\vspace{.5cm}
Institute for Theoretical Physics\\ University of Utrecht\\
Princetonplein 5 \\ 3584 CC  Utrecht\\ The Netherlands}\\[18mm]
 
{\bf Abstract}
\end{center}
The Born-Infeld theory of a toroidal D3-brane has an $SL(5,{\bf Z})$
U-duality symmetry. We investigate how this symmetry is reflected in the
supersymmetry algebra. We propose an action of the group on the gauge theory 
fields in the BPS sector by introducing an extra field together with an 
additional  symmetry,
and argue for the U-invariance of the degeneracies of the BPS spectrum.
\vfill
 
\pagebreak

\newsection{Introduction}
 
The BPS-sector of toroidally compactified type II string theories is well 
known to be symmetric under a large discrete symmetry group,  U-duality 
\cite{ht}. In fact, the half BPS particles 
(in the noncompact dimensions), arising from perturbative as well as
non-perturbative ten-dimensional states, transform in a single multiplet
of the U-duality group. Thus one can relate states such as wrapped D-branes, 
which
are non-perturbative from the string point of view, to states that have a
perturbative string description. In particular, U-duality predicts
the degeneracies of string states and dual D-brane states to be equal. This
prediction was checked for particular configurations by Vafa and Sen 
\cite{vafasen},
who used the gauge theory description of D-branes \cite{polchwitten}.
 
The gauge theory on a D-brane provides a description of not only the free 
brane itself,  
but also of bound states with lower dimensional branes and strings. These 
$1/4$ BPS states, preserving $8$ of the $32$ supersymmetries, 
are identified with configurations with non-zero electric and magnetic
fluxes \cite{tay}. For low enough dimensions, ($d\leq 4+1$) the flux quantum 
numbers of
the gauge theories are sufficient to describe all bound states with lower
dimensional objects than the brane itself, so one may
wonder how much of the U-duality remains valid in the field theory limit.
This question was addressed in \cite{piolobers, hacq},
partly motivated by the study of matrix theory \cite{bfss}. Although the
Yang-Mills Hamiltonian is not U-invariant, it was found in \cite{hacq}
that, remarkably, the part of the BPS-spectrum containing single BPS-states
(as opposed to those consisting of multiple states) exhibits U-invariant
degeneracies.

The gauge theory on the brane arises as an effective theory describing
the dynamics of the open strings glued to the D-brane. To first   
approximation this theory is indeed a Yang-Mills theory. However, when  
one wants to incorporate fields with large field strength compared to the
string scale, higher derivative terms begin to play a role. It was shown
by Leigh \cite{leigh} that the $\alpha'$ expansion sums to a Born-Infeld
gauge theory. (The extension of this theory to 
non-abelian gauge groups is still not fully
understood; for proposals and discussion see \cite{tseytlin,brecher}). It 
turns out that the U-symmetry that
was not fully realised by the Yang-Mills theory is more apparent in the
Born-Infeld theory. In \cite{hvz} the BPS masses of the compactified
D4-brane BI-theory were computed, and shown to be consistent with the
full $SO(5,5,{\bf Z})$ symmetry. In particular, the rank of the group,   
i.e. the number of D4-branes, appears on an equal footing with the
other quantum numbers. 

In this paper we will investigate the action of U-duality on the 
supersymmetric version of the Born-Infeld theory (as presented in 
\cite{aps}), specialised to four dimensions. The intention is to generalise 
the method and 
results of \cite{hacq}, where the four-dimensional Yang-Mills theory was 
investigated, to the Born-Infeld theory. The expectation is that, due to the 
more symmetric form of BI theory compared to Yang-Mills theory, we may 
find better agreement with the U-duality (in this case $SL(5,{\bf 
Z})$) predictions from string theory 
(in particular, we hope to also incorporate multiple BPS states in a 
U-dual degeneracy formula). 

We derive the (U-invariant) superalgebra and 
the BPS-equations (which of course coincide with the result in 
the bosonic theory). We find that we can express the equations naturally 
in terms of a field in the antisymmetric representation of $SL(5)$, whose 
zero modes are the quantum numbers. In identifying this field with the 
gauge fields we still have one extra local symmetry, which we may 
use to set the rank to a constant. In the context of classical solutions 
to the BPS equations, this allows us to indeed use U-duality as a 
solution generating technique. 

In order to study the degeneracies of the quantum system, we follow the 
philosophy of \cite{dvv,hacq} in quantising the theory restricted to the 
space of supersymmetric configurations. In a simple case the quantisation 
is very similar to that in the Yang-Mills case, where the BPS sector was 
shown to be related to a matrix string theory. We 
propose an action of the U-duality on the quantum theory, where we 
interpret the extra local symmetry just mentioned as a conformal symmetry, 
and the fixing of 
the rank to a constant as an analogue of light cone gauge fixing. 
We discuss multiple BPS states and present a U-duality 
invariant degeneracy formula for arbitrary states, depending only on two 
invariants constructed from the quantum numbers.

Finally we make some speculative comments on the similarity of the 
structure of the four-dimensional Born-Infeld theory to the self-dual 
tensor theory living on an M5-brane.

\newsection{$SL(5)$-invariance of the Hamiltonian and the BPS masses} 

\label{sectionbosonic}
The type IIB three brane compactified on a torus can be described by the 
Born-Infeld gauge theory. Configurations of this gauge theory on a three 
torus can be labelled by various integer quantum numbers, namely electric 
and magnetic fluxes, and momenta around the torus. These gauge theory 
quantities reflect the various bound states that the three brane in the 
string theory can be in. Electric fluxes correspond to bound states 
with strings, while magnetic fluxes, via electro-magnetic duality, 
represent D-strings. The world brane momenta are of course just the 
momentum numbers of perturbative string theory on a torus. Finally, the 
rank of the gauge theory reflects the number of three branes. The 
Born-Infeld action is only known for the $U(1)$ theory. 

Configurations in string theory on a three torus enjoy a U-duality 
symmetry $SL(5, {\bf Z})$, which acts on the various quantum numbers. 
The aim of this paper is to investigate the extent to which this symmetry 
is reflected in the gauge theory on the brane. In 
this section we will first analyse the form of the Hamiltonian of the 
(abelian) Born-Infeld theory, and demonstrate that the associated BPS 
mass, the minimal mass for a state with given quantum numbers, is 
compatible with the action of U-duality.
 
We begin with a derivation of the
bosonic D3-brane Hamiltonian along the lines of \cite{hvz}, and write it 
in a form that suggests an action of $SL(5)$ on the various fields. For 
simplicity, we
will set the various background tensor-fields to zero. The action is then
\be
S_{BI} = \int\! d^3xdt\,
            \frac{-1}{g_s}\sqrt{\det\bigl(-G-F\bigr)}.
\ee
$G$ is the metric on the toroidal three brane, in string units; $g_s$ 
is the ten-dimensional string coupling.
 
The $SL(5,{\bf Z})$ will act on the zero modes of the following ten fields,
\ba
\mbox{electric fields} &:& E^i = \frac{\delta\CL}{\delta F_{0i}},\nn\\   
\mbox{magnetic fields} &:& \ep_{ijk}B^k= F_{ij} = \partial_{[i}A_{j]},\nn\\
\mbox{momenta}         &:& P_i = F_{ij}E^j,\\
\mbox{rank}            &:& N=1.\nn
\ea
The zero-modes of these fields are the integer quantum numbers, and will be
denoted by lower-case letters.
 
The Hamiltonian squared can be derived to be, in terms of these fields,
\ba
{\cal H}^2 &=& \frac{\det G}{{g_s}^2}(N^2 +
G^{ik}G^{jl}F_{ij}F_{kl})\nonumber
\\&& + \left(E, P\right) \left(\begin{array}{cc} G& 0\\   
0& G^{-1}\end{array} \right)\left(\begin{array}{c} E\\ P\end{array}
\right).
\ea
From the string theory point of view, the first line gives the
contributions from D3 and D1-branes, while the last line contains the
string winding and momentum modes.
 
To bring out the $SL(5)$-symmetry we organise the fields in an
antisymmetric tensor,
\be
M = \left(\begin{array}{ccccc}
0& P_3 &-P_2& E^1 &F_{23}\\
-P_3 & 0 & P_1 &E^2 & F_{31}\\
P_2 & -P_1 & 0 & E^3 & F_{12}\\
-E^1&-E^2 & -E^3 & 0 & 1\\
-F_{23}&- F_{31}& -F_{12} & -1&0
\end{array}\right).
\ee
We also define the analogous matrix of zero-modes $m_{ij}$, with as 
entries the zero-modes of the fields.
 
The Hamiltonian squared can be 
expressed as 
\be
{\cal H}^2 = -\frac{1}{2}\mbox{Tr}{\cal G} M {\cal G} M,
\ee
with
\be
\label{moduli}
{\cal G} = \left(\begin{array}{ccc} \frac{G}{\sqrt{\det G}} &0&0\\
                                      0& {\cal G}_{44}&0\\
                                      0&0& {\cal G}_{55}\end{array}\right),
\ee
and ${\cal G}_{44}=\sqrt{\det G}$, ${\cal G}_{55} = \sqrt{\det
G}/{g_s}^2$. Non-zero off-diagonal entries would correspond to 
expectation values of the anti-symmetric tensor background fields.
 
$SL(5)$ acts by conjugation on the fields $M$ and the
background ${\cal G}$, leaving ${\cal H}$ invariant. The $SL(3)$ subgroup
residing in the top-lefthand corner is of course just the structure group
of the three-torus, leaving the coupling $g_s$ fixed and acting on
the momenta $P_i$, electric fields $E^i$ and magnetic fields
$B^i=\fr12\ep^{ijk}F_{jk}$ as vectors. The $SL(2)$ subgroup acting on the
$4,5$-indices is the familiar S-duality group, exchanging electric and
magnetic fields and inverting the coupling. It also acts on $\det G$,
when expressed in string units; as explained in \cite{hvz}, one should 
express the moduli in seven-dimensional Planck units to have a truly 
duality invariant formula. These two subgroups both leave
the rank $N$ fixed. The rest of $SL(5)$ mixes the rank with the fields,
and strictly speaking we only derived the Hamiltonian for the case
$N=1$, for lack of knowledge of the non-abelian version of the
Born-Infeld theory. However, based on the information we have from
string theory, we should expect the $SL(5)$ structure of the 
Hamiltonian   
as given above to be correct. In the gauge theory, the 
transformations
involving the rank should have an interpretation as 
Nahm transformations
\cite{hacq}.
 
The matrix $M_{ij}$ contains combinations of Born-Infeld fields that are 
not all independent. In particular, the momentum densities $P_i$ are 
the components of the Poynting vector, the exterior product of the 
electric and magnetic fields. This seems to spoil the $SL(5)$ symmetric 
structure. Remarkably, we can remedy this by viewing the entries 
of $M$ as a priori independent fields, and in addition expressing 
the relations between them by imposing an $SL(5)$ covariant vector $K^i$ 
of constraints. We define this $K$ using a five-dimensional $\epsilon$ 
symbol, defined via the metric ${\cal G}$, as 
\be
K^i=\fr18 \ep^{ijklm}M_{jk}M_{lm}\equiv *(M\wedge M)^i,
\ee
so that in terms of the Born-Infeld fields
\be
K=(NP_i-(E\wedge B)_i, -P\cdot B, P\cdot E).
\ee 
For the case that $N=1$, the first three components of the constraint 
$K=0$ give the definition of the momenta $P_i$ in terms of the 
electro-magnetic fields; the last two components are then automatically 
zero as well.  

Using the constraint $K=0$ we may derive the BPS mass formula for a given 
set of charges via the Bogomolny argument developed in 
\cite{hvz}. We may write, for any unit five-vector $\hat{v}^i$,
\be
\mbox{Tr}{\cal G} M {\cal G} M = \fr12 (M_{ij} + 
\fr12\ep_{ijklm}\hat{v}^k M^{lm})(M^{ji} + 
\fr12\ep^{jik'l'm'}\hat{v}_{k'}M_{l'm'}), 
\ee
(where we use ${\cal G}$ to raise and lower indices),
and split $M_{ij}$ into one part containing only 
the zero modes $m_{ij}$ and another with the fluctuations, $M'_{ij}$. 
From this we can, for any given $\hat{v}^i$,  minimise the energy, 
and maximising this minimum over $\hat{v}^i$ we find  the BPS 
bound 
\be 
M_{BPS}^2 = -\fr12 \mbox{Tr}{\cal G} m {\cal G} m + 2\sqrt{k\cdot k}
\label{bpsmass}
\ee
where $k= *(m\wedge m) = (p_i-(e\wedge b)_i, 
-p\cdot b,
p\cdot e)$ (recall that the lower case quantities are the zero modes of 
the corresponding fields). Note that the first term in (\ref{bpsmass}) is 
precisely the expression for ${\cal H}^2$ with the zero modes of the 
fields inserted; we see therefore that for configurations having non-zero 
five-vector $k^i$ (the $1/4$ BPS states, as opposed to the $1/2$ BPS 
states that have $k^i=0$), the mass is strictly greater than the zero mode 
value.
Hence, for such configurations the mass necessarily gets contributions  
from fluctuations in the fields.

This lower bound on the mass is only attained when the fields satisfy the 
BPS equations. These follow from the arguments leading to the BPS mass 
formula, and in this case require an expression involving the 
fluctuations in 
the fields to be proportional to the same expression with the zero modes:
\be
\label{bosonicbpsequation}
(|k|M' +  k*M') \sim   (|k|m + k*m). 
\ee
The proportionality factor is itself a function of space time.
In the following we will want to interpret this rather peculiar 
BPS equation, and, in particular, since the rank $N$ is part of the 
matrix $M$, to understand how to deal with possible fluctuations 
in $N$. In order to address these questions we will first rederive the 
above results from the supersymmetric Born-Infeld action, where the BPS 
equation follows from the requirement of unbroken symmetry. Then we will 
explore the solutions to the equations, using U-duality as a solution 
generating transformation. Here we will in particular demonstrate the 
consequences of transformations that formally give the rank $N$ 
fluctuating contributions. Finally we will propose a method of quantisation 
of the theory restricted to its BPS space, motivated by the $SL(5)$ 
symmetry, and demonstrate the U-invariance of the BPS degeneracies.   
First, however, we review the situation found by Hacquebord and Verlinde 
in their investigation of the theory in the Yang-Mills limit.

\newsection{Review of the Yang-Mills case} \label{sectionYM}
The question of degeneracies 
of BPS states of the three-brane was 
first investigated in the context of ${\cal N}=4$ super Yang-Mills theory on 
the torus \cite{hacq}. 
We will here first review this analysis, and then try to 
apply a similar method to the Born-Infeld case. 

The $U(N)$ super Yang-Mills theory is the reduction of ten-dimensional 
super Yang-Mills theory to four dimensions. Its bosonic fields are the 
vector potential, $A_\mu$, and six scalar fields $X^i$.  The theory has 
eight four-dimensional supersymmetries, four of which ($Q_1$) are the 
generators of the 
supersymmetries that are broken by the introduction of a brane, and act 
as shifts on the associated goldstone fermions $\lambda$. The other four 
generators ($Q_2$) are the unbroken supersymmetries, acting linearly. 

In ten-dimensional notation, the two charges are given by
\be
Q_1=\int \Tr \la, \quad Q_2=\int \Tr 
(E^i\Ga{0i} +\Ga{ij}F_{ij})\la. 
\ee
Together they form a superalgebra
\be
\{ \ol{Q}_a, Q_b\}=\bmat N\Ga{0} & 
2e_i\Ga{i}-\Ga{0ij}f_{ij}\\
2e_i\Ga{i}+\Ga{0ij}f_{ij} & 
2\Ga{0}p^0 +2\Ga{i}p^i\emat, 
\ee 
($a,b=1,2$) where lower case signifies the zero modes.

M(atrix) theory is the proposal to describe M-theory in its infinite 
momentum frame by means of the dynamics of zero-branes \cite{bfss}. From 
this perspective, Yang-Mills theory on a three-torus is a description of 
M-theory on a three-torus. In \cite{bss} it was demonstrated that the 
Yang-Mills theory superalgebra is indeed equal to the M-theory algebra in 
the light cone limit, $N\to\infty$. 

BPS states are states that are annihilated by certain combinations of the 
supercharges, depending on the quantum numbers. This implies that the 
variation of the fermions $\la$ should vanish under such symmetries, from 
which one can derive equations that the bosonic fields should satisfy, 
the BPS equations \cite{ow}. These were determined in \cite{hacq} to be
\be
E_i'\kappa^i=0, \quad E'_{[r}\kappa_{s]}=F'_{rs}.
\ee
The primes indicate that these equations only involve the non-zero modes 
of the fields; the vector $\kappa$ is the three dimensional part of the 
$SL(5)$ vector $k$ (defined below equation (\ref{bpsmass})), representing 
the momentum minus its contribution from the zero modes of the fields:
\be
\kappa= p-\frac{e\wedge b}{N}.
\ee 
In a suitable gauge the BPS equations become equivalent to the free field 
equations of a chiral two-dimensional model, with no dependence on the 
directions orthogonal to the vector $\kappa$. If we 
choose coordinates such that the momentum points in the 1-direction, 
\be
\partial_0A_i=\partial_1A_i,\quad \partial_0 X_I=\partial_1X_I,
\ee
with $i=2,3$, $I=1\ldots 6$. Furthermore, the $U(N)$ valued fields are 
required to 
mutually commute, but they are allowed to be identified periodically along 
the spacelike circle modulo a permutation. From this it was argued in 
\cite{hacq} 
that in the case of all electric and magnetic fluxes equal to 
zero, the solutions are characterised by the sigma model on 
\be
\frac{({\bf R}^6\times{\bf T}^2)^N}{S_N},
\ee
where $S_N$ is the Weyl group of $U(N)$. The vector field is responsible 
for the ${\bf T}^2$, whereas the six scalars parametrise ${\bf R}^6$. The 
model therefore reduces to a 
matrix string theory \cite{matrixstring}; its Hilbert space consists of all 
products of 
Hilbert spaces of strings of length $n_i$, such that their total length 
is $N$. In particular, \cite{hacq} considered the sector involving one 
long string, where the $S_N$ twists on the $A,X$ fields are such that 
they are described by fields taking values on a circle whose length is 
precisely $N$ times that of the physical circle. Quantising this system, 
it was found that the number of states of quantum numbers $N$, $p_1$ 
equaled $d(Np_1)$, defined by
\be
\sum d(n)q^n= 256 \prod \left( \frac{1+q^n}{1-q^n}\right)^8.
\ee
The inclusion of electric or magnetic 
fluxes in the 1-direction was argued to effectively decrease the rank to 
$N'=\mbox{gcd}(N,e_1,b_1)$. The reason is that in a $U(N)$ gauge 
theory a magnetic flux $b_1$ is realised by having twisted boundary 
conditions for the $SU(N)$ part of the gauge fields along the $2$ and $3$ 
directions. These twists should be two $SU(N)$ transformations which 
commute up to  a ${\bf Z}_N$ phase factor determined by the value of 
the magnetic flux modulo $N$. The point is that the $S_N$ twists making 
the string in a long string should commute with these magnetic twists, in 
order not to introduce an additional magnetic flux in the $2$ or $3$ 
directions.
This turns out to imply that the relevant permutation group can be maximally 
$S_{N'}$, thus effectively reducing the length of the single long 
string.

Therefore in this case the oscillation number 
of the long string adds up to $N'\times (p_1- (e\wedge b)_1/N)$, which is 
precisely the greatest common divisor of the five-vector $k$. The 
resulting degeneracy is then precisely that of (single 
particle) string states, and in accordance with U-duality it only depends 
on the U-invariant quantity $|k|=\mbox{gcd}(k)$.

In the long string sector, Hacquebord and Verlinde therefore find a 
degeneracy of states in agreement with predictions by U-duality. Sectors 
consisting of several shorter strings, however, do not respect the 
symmetry; this is related to the fact that the Yang-Mills Hamiltonian is 
not invariant under $SL(5)$. We will see below that using the Born-Infeld 
theory, which does have a U-invariant Hamiltonian, improves this situation. 

\newsection{Supersymmetric Born-Infeld theory}

We now want to apply the same methods to study the BPS spectrum of the 
Born-Infeld theory. First we will use the supersymmetric generalisation of 
the theory, described in \cite{aps}, to obtain expressions for the 
supercharges and their algebra. Remarkably, this algebra will turn out to 
be isomorphic to the superalgebra of M-theory compactified on a 
four-torus. This establishes that there exists an action of $SL(5)$, 
which is the associated U-duality. Then we derive, from the requirement 
that the superalgebra should have eigenvalues equal to zero, the BPS mass 
found previously from the bosonic argument.
 
The supersymmetric Born-Infeld theory in ten dimensions was obtained in 
\cite{aps} in the simple form
\be
\label{SBI}
S_{SBI}= -\int d^{10}x \sqrt{-\det ( \eta_{\mu\nu}+ F_{\mu\nu} - 
2\labar\Ga\mu\pd\nu\la + \labar\Gamma^\rho
\pd\mu\la\labar\Ga\rho\pd\nu\la )}.
\ee
(The background has been further simplified to $g_s=1, G=\eta$.)
This result was derived from a Green-Schwarz type action with manifest
target space covariance, by fixing the local $\kappa$-symmetry and going
to static gauge. Half of the $32$ supersymmetries of the target space are
broken; these are realised non-linearly on the action (\ref{SBI}). The
fermions $\la$ are the associated Goldstone fermions. The other sixteen
supersymmetries act linearly. The transformations can be split into
three parts: the simple supertranslations of the covariant theory, plus
the complicated kappa-symmetry and coordinate transformations needed to
restore the gauge. These add up to the rules \cite{aps}
\ba
\label{susytrans}
\del \labar &=& \epbar_1+\epbar_2\zeta + (\epbar_2\zeta -
\epbar_1)\Ga\mu\la\pd\mu\labar\nn\\
\del A_\mu  &=& (\epbar_2\zeta -
\epbar_1)\Ga\mu\la + (\epbar_2\zeta -
\epbar_1)\Gamma^\rho\la\pd\rho A_\mu + \pd\mu(\epbar_2\zeta -    
\epbar_1)\Gamma^\rho\la A_\rho.
\ea
$\zeta$ is a particular expression involving wedge products of the field
strength $F$ contracted with gamma-matrices; its specific form in the 
case $d=4$, putting all derivatives of the six scalars to zero, is given 
below. $\ep_1$ corresponds 
to the broken supersymmetry, the unbroken one is associated to a
combination of $\ep_1$ and $\ep_2$.
 
These expressions are valid in ten dimensions. The four-dimensional 
Born-Infeld theory we need can be easily obtained from this one by
dimensional reduction.
 
Unfortunately, the action (\ref{SBI}) and the associated transformation
rules are those applying to a $U(1)$ theory only. We will henceforth do
computations in this abelian theory, although at some points the rank
$N$ will be written to bring out the symmetry more clearly. In the end we
will make
some remarks on the alterations that should occur when non-abelian
theories are considered.

The programme now is to compute the supersymmetry algebra, and find
bosonic configurations which are annihilated by half of the
supersymmetries. To facilitate the computations, we will expand the
action to quadratic order in the fermions $\la$; this is enough to find
the purely bosonic central charges.
 
From the variations given in (\ref{susytrans}) it is straightforward to 
compute the two supercharge densities. We find, 
to linear order in the fermions, 
\be
J_1 = -2(-P^0 \Ga{0}\la + (E^i+P^i)\Ga{i}\la ),\quad J_2 = \zeta J_1.
\ee
In $d=3+1$, we can evaluate the gauge field dependent quantity $\zeta$ to be
\be
\zeta = \frac{1}{1+ B_{i}^2}\left(P^0 + (E_i(1+B^2)-E\cdot B B_i)\Ga{0i}
-\fr12\ep^{ijk}B_k\Ga{ij} + E\cdot B\Ga{0123}\right).
\ee
Inserting this in the expression for $J_2$ we find
\be
J_2 = 2(\Ga{0} -\fr12 F_{ij}\Ga{0ij})\la.
\ee
We can then use the Dirac brackets
\ba
&&\{-P^0 \Ga{0}\la + (E^i+P^i)\Ga{i}\la, \labar\}=\fr12, \nn\\
&&
\{-P^0 \Ga{0}\la +
(E^i+P^i)\Ga{i}\la,A_j\}=\fr12\{-P^0\Ga{0} + (E^i+P^i)\Ga{i}, A_j\} \la
\ea
to calculate the superalgebra
\be
\label{superalgebra}
\{\ol{Q}_a, Q_b\} = \bmat -2(p^0\Ga{0}+(p+e)^i\Ga{i}) & 2\Ga{0} +
f_{ij}\Ga{0ij}\\
2\Ga{0}-f_{ij}\Ga{0ij} & -2(p^0\Ga{0} +2(p-e)^i\Ga{i}) \emat,
\ee
($a,b$ take values $1,2$).
Recall that the lower case letters indicate the zero-modes of the
fields; here they appear, since the charges $Q$ are the integrals of the
densities $J$. If one takes the limit for infinite rank, $n\to\infty$, 
one would recover the Yang-Mills superalgebra (with a redefinition of 
$Q_{1,2}$). In this sense, we can view the Born-Infeld theory as the 
generalisation of Yang-Mills M(atrix) theory to finite rank $n$, where 
the light cone direction is taken to be finite as well.  

The superalgebra (\ref{superalgebra}) is closely related to the algebra
of 11-dimensional supergravity compactified on a four-torus. Concretely,
if we define the 32 component supercharge
\be
{\cal Q} = \bvec \Ga{0}Q_1\\ Q_2\evec,
\ee
and eleven-dimensional gamma-matrices
\be
\gamma_i = \Ga{i}\otimes\sigma_1,\quad \gamma_{11} = 1\otimes\sigma_3,
\ee
the algebra becomes
\be
\{\ol{{\cal Q}},{\cal Q}\}= -2 (p^0\gamma_0 + e^i\gamma_i + n\gamma_{11}
+ p^i\gamma_{i11} + \fr12 f_{ij}\gamma_{ij}),
\ee
which is precisely the eleven-dimensional supersymmetry algebra  
(compactified on a four-torus), with
$e_i$ and $n$ the compact momenta, and $p_i, f_{ij}$ the components of
the central charge (in the compact directions). (The central charge 
associated to the five-brane cannot take a finite value on a 
four-torus). This algebra is 
well known to enjoy the U-duality symmetry, $SL(5)$ in the case of a
four-torus compactification. Hence we can conclude that the BPS-mass
deduced from (\ref{superalgebra}) is invariant under $SL(5)$. To
compute it, we have to demand the matrix (\ref{superalgebra}) to have
eigenvalues equal to zero; this gives an equation for $p_0$.
For a simple case, say $p_1, n\neq 0$ (and the rest of the quantum
numbers zero), we find
\be
p_0^2 = (n + p_1)^2.
\ee
By acting with $SL(5,{\bf Z})$ we can transform to the case of
general quantum numbers, to find
\be
p_0^2 = n^2 + p_i^2 + e_i^2 + b_i^2 + 2\sqrt{(np_i- f_{ij}e_j)^2 +
(p_i e_i)^2 + (p_i b_i)^2}.
\ee
Recall that strictly speaking we have only demonstrated this for
configurations having $n=1$. This expression for the energy of a BPS 
state coincides with the one found via the Bogomolny argument in the 
bosonic theory (equation \ref{bpsmass}), where we put the moduli 
matrix ${\cal G}$ equal to the identity matrix.

\newsection{BPS-equations}
   
So far we found the U-invariant BPS mass spectrum from the 
supersymmetric Born-Infeld
action. In this section we will go on to compute the BPS equations that 
configurations saturating this bound have to satisfy. They will of course 
ultimately turn out to be equivalent to those found in section 
\ref{sectionbosonic}. The strategy to find the equations is to demand 
a purely bosonic state to have the variations of the fermions equal to zero, 
for supersymmetry parameter $\ep$ satisfying
\be
\label{susyep}
\{ \ol{Q},Q\}\ep =0.
\ee
Looking at equations (\ref{susytrans}) this gives (to linear order in the 
fermions) the equation for $\zeta$: 
\be
\epbar_1 +\epbar_2\zeta = 0,
\ee
or equivalently,
\be
{\cal Z}\ep=\bmat 1 &\zeta \\ \ol{\zeta} & 1\emat \ep = 0
\ee
Here $\ol{\zeta} = -\Ga{0}\zeta^\dag\Ga{0}$, and there is a relation
$\ol\zeta\zeta=\zeta\ol\zeta=1$ \cite{aps}. 
 
We have to insert the general form of $\ep$ satisfying (\ref{susyep}) 
(given the central charges) in this equation, and then solve for $\zeta$, 
i.e. determine the form of the gauge fields. For general quantum numbers, 
this seems to be rather complicated, until we notice that in the abelian 
case ($N=1$) we may write
\be
\label{MZ}
(N^2 + B_i^2){\cal Z}= M^2,
\ee
with the matrix $M$ defined as
\be
\label{defM}
M= \bmat -2(P^0\Ga{0}+(P+E)^i\Ga{i}) & 2N\Ga{0} +
F_{ij}\Ga{0ij}\\
2N\Ga{0}-F_{ij}\Ga{0ij} & -2(P^0\Ga{0} +(P-E)^i\Ga{i}) \emat.
\ee
Remarkably, this matrix is identical to the supersymmetry algebra
(\ref{superalgebra}), except that we have put in the complete fields for
the
zero-modes. We know how $M$ transforms under the U-duality group, namely
similarly as the supersymmetry algebra, which we will call $m$. (The 
transformation of 
$M^2$, and hence of $\zeta$, on the other hand is rather complicated).

The BPS-equation is therefore found by demanding
$M^2\ep = 0$
for an $\ep$ annihilated by the zero mode $m$ of $M$; one can easily 
verify that this may be simplified to 
\be
M\ep =0.
\ee 
To extract the precise BPS condition, we write the condition 
$m\ep=0$ in a convenient form as follows. Note that
\be
\fr18  m\sigma_3m\sigma_3 = |k| 1\otimes 1 - (np_i
-f_{ij}e_j)\Ga{0i}\otimes\sigma_1 - p\cdot b \Ga{0123}\otimes i\sigma_2 +
p\cdot e  1\otimes \sigma_3
\ee
where $k$ is the five-vector whose components are the coefficients of the
other terms. This is precisely the same vector $k$ that we encountered 
before. The expression is a projection operator (times a constant), it 
can be written as 
\be
\fr18 m\sigma_3m\sigma_3 = |k| + k^i\tilde\gamma_{i},
\ee
for suitable $\tilde\gamma_i$ satisfying a five dimensional Clifford
algebra. From this we conclude that the condition $m\ep=0$ can
be expressed as
\be
\ep = \sigma_3m\sigma_3(|k| - k^i\tilde\gamma_i)\xi,
\ee
for arbitrary spinor $\xi$. Hence the BPS-equation becomes
\be
M\sigma_3m\sigma_3(|k| - k^i\tilde\gamma_i)=0.
\ee
If we work this out, for general charges, this yields the set of conditions
\ba
&&p_0(|k|M_{ij} + \ep_{ijklm}M_{kl}k_m)-P^0(|k|m_{ij} +
\ep_{ijklm}m_{kl}k_m) = 0\\
&&M_{ij}(|k|m_{ij} + \ep_{ijklm}m_{kl}k_m) - P^0p^0|k| = 0\\
&&\ep_{ijklm}M_{jk}(|k|m_{lm} + \ep_{lmabc}m_{ab}k_c)- P^0p^0k^i=0\\
&& k_iM_{ij}=0.
\ea
We naturally identified the matrix $M$ acting on spinors as (\ref{defM}) 
with the five by five antisymmetric matrix $M_{ij}$, and likewise for the 
zero mode $m_{ij}$. Apart from the matrix $M$ and its zero modes $m$, there 
also appears 
the Hamiltonian, $(P^0)^2=\fr12M_{ij}M_{ij}$, and its zero mode 
$(p^0)^2=\fr12 m_{ij}m_{ij} + 2|k|$. It can easily be checked 
that the last three equations all follow from the first one (one also 
has to use that $K\equiv M\wedge M=0$), so that ultimately the BPS-equation 
just becomes 
\be
\label{bpsequation}
p_0(|k|M_{ij} + \ep_{ijklm}M_{kl}k_m)-P^0(|k|m_{ij} +
\ep_{ijklm}m_{kl}k_m) = 0.
\ee
This is precisely the statement (\ref{bosonicbpsequation}) obtained from the 
bosonic theory. Here, however, we immediately also get the factor of 
proportionality, $(P^0-p_0)/p_0$. 

To recapitulate, we have reproduced the BPS-bound and the BPS-equation 
from the supersymmetric theory. Now we would like to study solutions to 
this equation and interpret the meaning of space-dependent $M_{45}$, the 
rank of the gauge group. We will deal with this by noting that the 
definition of $M$ from the relevant quantity ${\cal Z}$ leaves room for 
an extra rescaling by a fluctuating factor, which we will use to put 
$M_{45}$ to a constant. Then we can use 
$SL(5)$ as a solution generating transformation in the classical theory. 
After that we will study the quantum properties, by imposing the BPS 
equation on operators, and try to determine degeneracies of states of 
given quantum numbers $m_{ij}$.

\newsection{Classical solutions of the BPS equations} 
 
We will now study solutions to the Born-Infeld BPS equations. Ideally, 
one would like to solve the BPS equations for a simple configuration, and 
then invoke the duality to generate the solutions at different quantum 
numbers. This will be indeed the strategy we will follow. There are two 
questions that need to be resolved for this to work, however. Firstly, in 
fact we only know the Born-Infeld theory for the abelian case. We will 
make the assumption that the non abelian generalisation exists and shares 
the features of the Yang-Mills theory in its BPS space discussed in 
section (\ref{sectionYM}), that the fields effectively diagonalise and 
can be described as abelian fields on a longer circle. The second problem 
is that a duality transformation acting on the matrix $M$ will in general 
not keep the component related to the rank a constant number. To solve 
this we will argue that $M$ is not quite the gauge theory quantity, but is 
only related to it up to a local scale transformation, which may be used to 
`gauge' the rank to a constant.

To start with we will solve the BPS equation (\ref{bpsequation}) in the 
simplest possible case, where the only non zero quantum numbers are $n$ 
and $p_1$; we will also first take $n=1$ and postpone the discussion on 
the non abelian case.

We have, with $p^0= n+p$,
\be
P^0=N+P^1,\quad E^1=B^1=0,\quad E^{i}=F^{1i},\quad P^i=0,
\label{soln}
\ee
where $i=2,3$. Since, in this particular case, $E^i=F^{0i}$, we recover 
the same equations as those valid in the Yang-Mills situation. As was 
done there, we choose the gauge $A_0=A_1=0$, so that the gauge fields 
$A_{2,3}$ satisfy 
\be
(\partial_0-\partial_1 )A_i=0.
\ee
To obtain a genuine gauge theory we of course set $N=n$. The solutions to 
the BPS equations are then parametrised by two functions $E_2(t+x_1)$ 
and $E_3(t+x_1)$, satisfying 
\be
\int E_{2,3}=0,\quad \int E_2^2 +E_3^2 = np_1.
\ee
Note that in principle the BPS equation does not rule out dependence on 
the coordinates $x_2$ and $x_3$; this would, however, not be consistent 
with the Maxwell equations (although often BPS equations imply equations 
of motion, this is not in general the case). 

Now we will discuss the generalisation to the case that the rank $N>1$. 
The problem is that the form of the action of the non abelian Born-Infeld 
theory is yet unknown, even in the bosonic case. Tseytlin proposed a 
definition of the theory using a symmetrised trace   
prescription \cite{tseytlin}, and it was argued in \cite{brecher} that
this should be the only action allowing a supersymmetric extension. The 
precise form of this extension has not been found. We will make some 
assumptions on the alterations that arise, guided by the situation in the 
Yang-Mills theory, and requiring that the duality group have some 
sensible action on the fields, at least in the BPS sector.

Some of the formulas appearing in this paper seem to have a natural
generalisation to arbitrary $N$; in particular, one would expect the  
superalgebra to remain of the form presented in (\ref{superalgebra}),
with $N$ as the coefficient of the $\Ga{0}\otimes\sigma_1$ part, so that the
relation to the eleven-dimensional algebra be preserved.

The introduction of a matrix $M$ such as in the abelian case is now 
questionable, however. Merely inserting non-abelian matrices for the
entries does not give the desired squaring to something of the form of 
${\cal Z}$. Besides, it is
not quite clear what we would mean by this. Though electric and magnetic
fields are of course easily converted to Hermitian matrices, the
prescription for the momenta $P$ is already more complicated. One might  
for instance define it to be the anticommutator of the appropriate
$E$'s and $F$'s. The matrix we would naturally introduce for the rank $N$ 
is the $n\times n$ identity matrix, the trace of which of course is $n$. 
However, since duality mixes up the various components, these
relations are no longer preserved. What is more, since $N$ may acquire a
different value, the size of the component matrices must change as well;
it is hard to imagine a consistent action of the duality group on general
fields with this property.
 
To resolve these problems we make the following assumptions. We assume 
that the BPS configurations of the non abelian theory will behave like 
those of the Yang-Mills theory, in the sense that the fluctuations of 
the fields will be simultaneously diagonalisable and that they will all 
depend again on only one coordinate. Again, there may be twists in the 
theory making the period of this coordinate $n'$ times as 
long, where $n'=\mbox{gcd}(n,e_1,b_1)$. This makes the theory effectively 
an abelian one. We further assume that this abelian BPS sector satisfies 
the generalisations to larger $n$ of the equations presented in the 
original case $n=1$. In this way, the $SL(5)$ duality does have a 
reasonable action on the BPS sector of the theory. 

Using these assumptions, the solution calculated above can be 
straightforwardly generalised to higher rank $n$. The equations are the 
same, so we again have fields that are left moving. The coordinate 
however now has periodicity $n$ (or $n'$ in the presence of the fluxes as 
argued above). The integral of the fields over this 
space then incorporates both the trace and the integration over the real 
domain. We choose the function $N$ to be one, so that its integral indeed 
equals $n$. The other fields then again should have integrals as above.
  
So far we discussed only the BPS equations in the case of simple fluxes,
with only $p_1$ and $n$ unequal to zero. The equations were the same as
those obtained in the Yang-Mills theory. We now turn to more general
fluxes, where unlike the Yang-Mills case the BPS equations become more
complicated. We will argue that their solutions can be found using 
$SL(5)$ transformations.

$SL(5,{\bf Z})$ acts on the zero mode matrix $m_{ij}$ by conjugation. Any 
configuration with non-zero $k=m\wedge m$ can be mapped to one with only 
non-zero $n'$ and $p'_1$, such that $\mbox{gcd}(k^i)=n'p'_1$ and 
$n'=\mbox{gcd}(n, p_i, e_i, b_i)$ (see e.g. \cite{igusaconforto}). If the 
matrix $M$ transforms similarly the BPS equations are also covariant 
under this group, provided one also transforms the five by five moduli 
matrix ${\cal G}$ (\ref{moduli}) (off diagonal terms can also appear in 
this metric, and correspond to non-zero anti-symmetric tensor fields). An 
$M_{ij}$ (with zero modes $m_{ij}$) 
solving the equation, will therefore, when transformed to $M'$ (with zero 
modes $m'$) also solve the BPS equation at the transformed moduli (note 
that $P^0$ is invariant if we also transform the moduli). 

The new solution, however, will only correspond to a possible gauge 
theory solution if the new component $M_{45}$ is a constant, to be 
identified with the rank. 
This is certainly the case under transformations belonging to $SL(2,{\bf 
Z})\times SL(3,{\bf Z})$, which leave $N$ invariant, but not in general. 

The resolution is to recognise that the relation between the matrix $M$ 
and the gauge theory fields $E_i$, $B_i$, $P_i$ and 
$N$ is not unambiguous. The original BPS equation was
\be
{\cal Z}\ep=0,
\ee
and we reformulated this in terms of a matrix $M$ defined in equations 
(\ref{MZ},\ref{defM}). This reformulation is not unique however, as can 
be seen from the fact that $\zeta$, the top righthand block of ${\cal 
Z}$, is a function only of the quotients $M_{ij}/N$:
\ba
\zeta &=& \frac{1}{N^2+B^2}\Bigg[NP^0  + (\frac{E_i}{N}(N^2+B^2) 
-\frac{E\cdot B}{N}B_i)\Ga{0i} \nn\\&&- \fr12\ep_{ijk}P^0B_k\Ga{ij} + 
E\cdot B \Ga{0123} \Bigg],\nn\\
 &=& \frac{1}{1+(B/N)^2}\Bigg[P^0/N + (E_i/N(1 + (B/N)^2) -(E\cdot
B)B_i/N^3)\Ga{0i}\nn\\  && -\fr12 \ep_{ijk}P^0B_k/N^2 \Ga{ij} + E\cdot
B/N^2\Ga{0123}\Bigg].
\ea
Therefore, if we scale the matrix $M$, the physical fields present in 
$\zeta$ remain the same, demonstrating that $M$ only equals the various 
fields up to a factor which may vary over spacetime. 
Given a solution for $M$, we may therefore derive the corresponding gauge 
fields by rescaling this $M$ such that the entry in $M_{45}$ becomes the 
appropriate constant, and then reading off the values of the fields. 

The strategy to solve the BPS equations for general fluxes is then
to first transform the fluxes by an element $g\in SL(5,{\bf Z})$ to the 
simple case. Then we know the BPS equations; where formerly we put $N$ to 
a constant, now we demand that the combination that under $g^{-1}$ will 
be mapped to the $45$ component of $M$ is a constant. We further impose 
the equation
\be
K=NP_1- E_2B_3+ E_3B_2=0;
\ee
this leaves us with two arbitrary functions. On these we impose the 
condition that the zero modes of all fields are the correct ones. If we 
then transform back using $g^{-1}$, the resulting configuration will 
satisfy the BPS equation for these fluxes, will have the correct zero 
modes and will obey $N=n$. This guarantees that we have an allowed BPS 
configuration.    

As mentioned earlier, fields satisfying the BPS equations and also the 
condition $M\wedge M=0$ are not yet the complete story: we still have to 
demand them to satisfy the equations of motion and Bianchi identities. 
These will tell us how the fields depend on the coordinates. To show that 
in general the fluctuation parts of the $E$ and $B$ fields again are left 
movers, depending on the coordinate in the direction of the zero 
mode of the momentum of the fluctuating fields, we need the relation 
between $F_{0i}$ and $F_{ij}$ in the general case. The general expression 
for $F_{0i}$ in terms of the electric and magnetic fields is
\be
F_{0i}= (M^2)_{i5}/P^0.
\ee
We can now use the BPS equation (\ref{bpsequation}) to derive
\be
F_{0i}=\frac{1}{p^0}(m_{ij}+\epsilon_{ijklm}m^{kl}\hat{k}^m)M_{j5}.
\ee
We will analyse two situations, $k=k_1$ and $k=k_1+k_4$. For other 
configurations the discussion is similar.

For $k$ having only a component in the one-direction, we have that 
$M_{15}=0$. Since the matrix multiplying $M_{j5}$ is anti-symmetric, we 
see that $F_{02}$ is a linear combination of $B_3$ and $N$. If we 
restrict to fluctuations, we therefore have
\be
F'_{02}=\mbox{const}\cdot F'_{12},
\ee
and similarly for $F_{03}$.

For $k$ having also a component in the four-direction, we do not 
anymore have $B_1=0$; however, in this case $B_1$ is a constant, since 
from $k^iM_{ij}=0$ we see that it is proportional to $N$. For the 
fluctuations, the relations therefore remain as above.

This demonstrates that in general  we still require (fluctuating parts of) 
the fields $A_2$ and $A_3$ to satisfy  the equations of a chiral boson, 
be it that the `speed of light' now deviates from one. 

This ends the 
discussion of the classical gauge field configurations in the BPS sector of 
the theory. In the next section we will again study the BPS sector, but 
there we will try to quantise the theory, with the BPS requirements as 
constraints.

\newsection{Quantisation of the 
Born-Infeld BPS states}

We have seen that $SL(5,\bf{Z})$ is a symmetry of the BPS equations, 
and can be used to generate solutions. It is a different question 
whether the quantum theory will enjoy the symmetry, so that the duality 
group will be a symmetry of the space of states and the associated 
degeneracies.

The idea is not to quantise the complete Born-Infeld theory, but to 
consider only the phase space corresponding to configurations that are 
supersymmetric. This system is then to be quantised. The expectation is 
that, since supersymmetry is so robust under quantum corrections, this 
gives an adequate description of states and processes in the BPS sector. 

To investigate this question we have to represent the fluctuating 
electric and magnetic fields as operators on a Fock space. The 
requirements will be that $E$ and $B$ satisfy the appropriate 
commutation relations, and that all the fields will have the 
prescribed integer expectation values when acting on states with 
the associated quantum numbers. We will construct such operators 
using the symmetry group and demonstrate that they have the desired 
properties. 

In dealing with the rank $N$, we will use an analogue of the classical 
situation described before. There we saw that scaling all fields was a 
symmetry, so that $N$ could be put to the appropriate constant value. 
Here we will propose to identify this symmetry with a gauge symmetry, 
generated by the constraint $K=0$. In the BPS sector, the system 
effectively reduces to a string theory, where $K$ takes the role of the 
stress energy tensor. The choice $N$ is constant is then simply the 
familiar light cone gauge.  

We will start by quantising the simplest case, 
where only $n$ and $p_1$ are unequal to zero. The BPS equations are
\be
P_0=N+P_1,\quad E_1 = B_1 = 0,\quad E_2 =B_3,\quad  E_3=-B_2,
\ee
which in this case implies $F_{0i}=F_{1i}$ for $i=2,3$. The equations 
reduce the phase space to effectively only two independent 
functions, $E_2$ and $E_3$, (plus the six scalars), which can all be 
chosen to depend only on $t+x$. Canonical commutation relations imply:
\be
[E_{i}(x_1), E_{i}(x_2)]= i\fr12\partial^1 \delta (x_1-x_2) \mbox{  (for 
$i=2,3$)}. 
\ee
As we saw, the fields $E_i$ are   
functions of a coordinate whose periodicity is $n$ times as long as 
that of $x_1$. The two fields can then be quantised as 
\be
E_2=\sum_k \frac{1}{\sqrt{2n}}\alpha^2_\frac{k}{n} 
e^{i\frac{k}{n}(\tau+\sigma)},\quad 
E_3=\sum_k \frac{1}{\sqrt{2n}}\alpha^3_\frac{k}{n} 
e^{i\frac{k}{n}(\tau+\sigma)}, 
\ee
(and similarly for the scalars) where 
$[\alpha^i_\frac{k}{n},\alpha^j_\frac{l}{n}]=\frac{k}{n}\delta^{i,j}_{k,-l}$, 
and $\sigma$ runs from $0$ to $n$.
Then $P_1$ can be solved using the identity $NP_1-(E\wedge B)_1=0$ 
(with $N$ the identity matrix). Imposing that $\int P =p$ gives that 
the number of states is $d(np)$.

We propose to view the above computation as a string degeneracy 
computation in the light cone gauge in the following way. Let us 
assume an underlying theory where the components of $M$ are all 
functions of space time (including the $M_{45}$ component which 
represented the rank), and which satisfy commutation relations of the 
form
\be
[M_{ij}(x),M_{kl}(y)]=i\epsilon_{ijklm}\partial^m\delta (x-y).
\label{comrelM}
\ee
Formally the index on the derivative runs over two additional dimensions, 
$4,5$. We will, however, always demand that the BPS configuration 
only depends on time and the spatial coordinate in the direction 
of $p$ (or rather $\kappa$, the first three components of $k$ as 
introduced in section \ref{sectionYM}); all other derivatives are put to 
zero. Furthermore $M$ satisfies the constraint $K=M\wedge M=0$. 
The BPS conditions put various conditions on the components of $M$, and 
furthermore in the present case restrict all fields to depend only on
$x_1-t$.
We label the remaining independent fields $N=\partial X^+,\quad P_1 = 
\partial X^-, E_{2,3}=\fr{1}{\sqrt{2}}\partial X^{2,3}$, and the other six 
scalars $\frac{1}{\sqrt{2}}\partial X^i$. In this form the commutation 
relations 
(\ref{comrelM}) are precisely those of the left moving sector of a 
string, while the constraint $K$, which equals $K_1=\partial X^+\partial 
X^- 
- \fr12 \partial X^i\partial X^i$, takes the form of the associated stress 
energy tensor. We now wish to interpret the choice $N=\partial X^+=n$ 
as a gauge fixing of the symmetry generated by the local constraint $K$, 
analogous to the 
conventional light cone gauge. After this fixing, $P_1=\partial X^-$ 
is to be solved from the constraint, and we end up with precisely the 
same representation of the various fields, with the correct 
commutation relations, as above. We therefore can view this 
description in terms of left moving bosons plus a gauge symmetry as 
effectively providing a way of generating a representation of the 
fields $E_i$, $P_1$, with the correct commutation relations and zero 
modes. Furthermore, the identification of $K^1$ with the stress 
energy tensor $T$ explains the degeneracy formula, since we have a 
left moving string theory at level $L_0 = p^+p^--\fr12(p^i)^2 = np_1$.

It turns out that the generalisation to a theory with more degrees of 
freedom plus a gauge symmetry that we just introduced can be used 
also to quantise the theory for general quantum numbers. The strategy 
is to start with the previous case and, before fixing the gauge, 
acting on the fields with $SL(5,\bf{Z})$. This will transform $M$ to 
a new $M$ where all components are linear combinations of the various 
$\partial X$. In particular $M_{45}$ is now changed from $\partial 
X^+$ to a linear combination involving other $\partial X^i$ as well. 
Using the fact that the theory with ten bosons and supersymmetry is 
Lorentz invariant, we may equivalently fix the new $N$ to be equal to 
its zero mode $n$, and again solve for $P_i$; if we identify the 
gauge theory fields with the appropriate $\partial X$'s {\em in the 
gauge where $N$ is constant}, these
fields will again satisfy the appropriate commutation relations, and 
have the right zero-modes. Furthermore, the degeneracy is again 
computed in the conformal field theory at level $|k|$.

To verify these assertions we will need to consider various subgroups 
of the duality group. First of all, the groups $SL(2,{\bf Z})$ and 
$SL(3,{\bf Z})$, electro-magnetic duality and the symmetries of the 
torus, are already evident symmetries in the gauge theory 
description, without the introduction of the bosons, since they do 
not affect the rank $N$.  

We will first consider the subgroup $SL(4,{\bf Z})$ that leaves $k=k_1$
invariant. Inserting in the matrix $M$ for the case of non zero $n$ and
$p_1$ the various $\partial X$'s found above, by acting with the
transformation we obtain a new $M'$ which satisfies the new BPS equations.
Since solutions of the resulting BPS equations still depend only on $t$
and $x_1$ in some fixed combination, acting with the transformation leaves
the commutation relations (\ref{comrelM}) invariant. After the
transformation, the zero modes of $M'$ are precisely the $SL(4)$
transformed ones. In the previous case, we then picked a gauge where $X^+$
was proportional to the worldsheet time coordinate, so that $N=\partial
X^+$ was a constant. Using Lorentz covariance, however, we may
alternatively gauge fix any other combination of the $X$ instead of 
$X^+$, without affecting the theory.
Obviously we now pick the gauge where $N'=n'$. We may then again solve for
$P_1$ ($P_{2,3}$ remain zero), and obtain a representation for the fields
$E$ and $B$ satisfying the commutation relations. Since $K^1$ remains
invariant, the stress energy tensor remains the same, as well as the level
$np_1$, so that we find the same degeneracy. 

The case where $k$ is not left fixed is slightly more complicated; we 
will consider an $SL(2)$ transformation on the simple case, changing 
$k_1=np_1$ to 
$k'_1=rk_1,k'_4=sk_1$, with $r,s$ mutually prime. Before application of this 
$SL(2)$ transformation we denote $N=\partial X^+, P_1=\partial X^-$ and 
$E_{2,3}=\partial X^{2,3}$ as before, and furthermore $B_2=-E_3, 
B_3=E_2$. Acting on these fields with the element of $SL(2)$ results in 
a new configuration where $N, E_2$ and $E_3$ are $r$ times their former 
expression, whereas $P_1, B_{2,3}$ remain unaffected. Furthermore, $P_2, 
P_3$ and $B_1$ also get a value different from zero. 

If we would quantise the simple theory as before, and apply the 
transformation, we do of course get the desired degeneracy, $d(np_1)$. 
However, this 
is not quite correct, since the new configuration does not satisfy the 
required commutation relations: the new $E$ and $B$ commutator is $r$ 
times too large. This can also be understood from the transformation of 
the $M$-commutation relations, as given in equation (\ref{comrelM}): 
formally, the only derivative that was non-zero, $\partial^1$, is 
transformed into $r\partial^1 + s\partial^4$. The new configuration can 
be taken to depend only on $x_1$, and we demand $\partial^4$ to be zero. 
This leaves us with a commutation relation that is $r$ times its previous 
value. 

To ensure the correct commutation relations after the $SL(2)$ 
transformation, we therefore have to scale the commutation relations of 
the $\partial X^i$ down by a factor $r$. However, if we then compute the 
degeneracy in the pre-transformation, simple state with only $p_1$ and 
$n$ non-zero, with these rescaled relations we seem to get a degeneracy 
which is $d(rnp_1)$ rather than $d(np_1)$. 

The situation is saved by the observation of \cite{hacq}, that the 
allowed twists generating long strings take values only in $S_{N'}$, 
rather than $S_N$, where $N'=\mbox{gcd}(n,b_1)$. The fields after the 
transformation have a period which is $r$ times smaller than naively 
expected. Therefore, not all oscillators are allowed: the oscillator 
numbers are $r$ times larger than without this requirement. In the end, 
this combination of rescaling the commutation relations, plus rescaling 
the allowed periodicity of the fields, gives us again the correct result 
for the degeneracy: $d(\mbox{gcd}(k))=d(p\cdot\mbox{gcd}(n,b_1))$. 

Acting then on this configuration with $SL(4)$ keeping the vector $k$ 
invariant works similarly as in the case where $k$ was pointing only in 
the $1$-direction, and gives the same degeneracy. Other configurations of 
$k$ can be trivially obtained using $SL(2)\times SL(3)$ transformations. 

This concludes the demonstration of U-invariance of the degeneracies in 
the long string sector. Next we will investigate the contributions of 
multiple string states. 

\newsubsection{Multi-particle BPS states}

So far we considered the degeneracy associated to the sector involving 
`long strings', i.e. the configurations containing only one single BPS 
bound state. It is in principle also possible to construct multi-particle 
BPS states, whose charges add up to the total charge. In the symmetric 
product string sigma model these are identified with the states in those 
twisted sectors in the conjugacy class where $N$ is partitioned in more 
than one piece. For these states to be BPS, it is necessary that all 
component states are annihilated by the same supersymmetry, or, 
equivalently, that the total energy is precisely the sum of the 
constituent BPS energies. In  
M-theory, this implies that the charge matrices of the component states 
should be proportional. In Yang-Mills theory this turned out not to be 
the case \cite{hacq}; there was a greater freedom in the distribution of 
the charges over the constituents, in particular with respect to the 
value of the momentum zero mode. This was related to the fact that the 
Yang-Mills BPS mass was not symmetric under duality. In the present case, 
for generic moduli it is clear that all charges should align for the 
component masses to add up to precisely the composite BPS mass. This is 
of course also evident from the relation the supersymmetry parameters 
have to obey, $m\ep=0$. We can therefore conclude that in the Born-Infeld 
case the counting of degeneracies, including those related to 
composite BPS states, gives the result expected from M-theory.

To give an explicit expression for the total U-invariant degeneracy, we
first note that, since every component charge matrix ${m}^\alpha_{ij}$ is
proportional to the total charge $m_{ij}=\sum_\alpha m^\alpha_{ij}$, the
different compositions of $m_{ij}$ in sets of $m^\alpha$  coincide with all 
possible ways one can
partition the number $|m|\equiv \mbox{gcd}(m_{ij})$ in a collection of 
$n_\ell$ states of 
$|m^\alpha | = \ell$, so that $\sum \ell n_\ell = |m|$. (The long string has 
$n_{|m|}=1$, the other $n_\ell=0$.) Each short, component, string has its 
own vector $k^\alpha=m^\alpha\wedge m^\alpha$, with $|k^\alpha |\equiv 
\mbox{gcd}((k^\alpha)^i)$. The 
degeneracy associated to such a short string of length $\ell$ is therefore 
given by $d(|k^\alpha |)$. For the long string, we have $|k| = r |m|^2$, 
since
the vector $k$ is quadratic in the components of $m$. The number $r$ is a
fixed integer depending on the state. Therefore, we have $|k^\alpha |=
r\ell^2$, so that a string of length $\ell$ can be in $1/2 \cdot d(r\ell^2)$
different bosonic states, and equally many fermionic ones. The total
BPS-state is a symmetric combination of the partial ones, resulting in a
degeneracy formula 
\be \sum_{|m|} D(r,|m|) q^{|m|} =\prod_\ell
\left(\frac{1+q^\ell}{1-q^\ell}\right)^{\frac{1}{2}d(r\ell^2)}, 
\ee 
where as mentioned $r=|k|/|m|^2$. The total degeneracy therefore depends 
on the two U-invariants, $|k|$ and $|m|$.
\pagebreak

\newsection{Discussion} 
We analysed the Born-Infeld theory of a compactified D3-brane, and argued 
for a realisation of $SL(5,{\bf Z})$ U-duality on the $1/4$ BPS sector of 
the theory (making a number of assumptions on the structure of the 
non-abelian theory). The action on the quantum numbers was well 
known already, the surprising fact was that there seems to exist an 
action of the duality also at the level of fields. These local degrees of 
freedom, assembled in the matrix $M_{ij}(x)$, are not exactly the fields 
of the gauge theory. Rather, $M$ contains one more degree of freedom, 
which is compensated by a local symmetry acting on $M$. The 
Born-Infeld theory, 
in its BPS sector, was argued to be a gauge fixed version of this 
extended model: the gauge condition is that the rank is a constant.
On the extended model $SL(5)$ has a natural action; however, this action 
does not in general preserve the gauge. The action at the level of the 
Born-Infeld theory therefore has to be supplemented by a compensating 
transformation restoring the gauge. 

Furthermore, by interpreting the model (reduced to the BPS sector) as a 
string theory in light cone gauge, we computed the degeneracies of the BPS 
states, and found them to be in agreement with the predictions of 
string/M theory.

We conclude by making some speculative remarks on some striking 
similarities of the model presented 
here to the self-dual tensor theory, the low energy theory living on 
the M theory five-brane. We will consider this theory compactified on a 
five-torus. In a Hamiltonian approach \cite{hen}, one can 
describe the theory in terms of a spatial two-form $M_{ij}$ defined in 
terms of the self-dual field strength $F_{\mu\nu\rho}$ as
\be
M_{ij}=\frac{1}{6}\ep_{ijklm}F^{klm}.
\ee 
$M_{ij}$ of course depends on all six spacetime coordinates. The energy 
and momentum densities of the theory are
\be
P^0=M_{ij}^2,\quad P^i= *(M\wedge M),
\ee
which we recognise as our hamiltonian and $K$-vector, respectively. The 
Dirac brackets were calculated in \cite{hen} to be
\be
[M_{ij}(x),M_{kl}(y)]=\ep_{ijklm}\partial^m\delta (x-y).
\ee
This is of course the same as equation (\ref{comrelM}), but in this case 
all coordinates are truly five-dimensional. The $SL(5)$ 
symmetry is evidently the purely geometrical symmetry of the 
five-torus.

A supersymmetric version of the M5 brane (in the PST formulation 
\cite{PST}) was studied in \cite{ST}. The supersymmetry algebra was 
calculated, and, like in our Born-Infeld calculation, it was 
shown to be precisely the M theory algebra. A compactified five-brane (on 
a five torus) of course has more charges; apart from the ten fluxes of 
$M_{ij}$, one generically also has the five momenta $P^i$, and the number 
of five-branes $N$. According to the correspondence between the momenta 
and our vector $K$, it seems that to make the connection to the 
three-brane theory one should demand all momenta to vanish. If we also 
set the five-brane number $N$ to zero, the supersymmetry algebra becomes 
similar to the one of interest in this paper; furthermore, one finds the 
same BPS equation for $1/4$ BPS states!

It is tempting to interpret the condition that the momenta vanish as the 
analogue of $L_0-\ol{L}_0=0$ in string theory, and view the five-brane 
theory as an underlying theory for the gauge theory. Interestingly, the 
rank of the gauge theory would then be just one of the fluxes of the 
compactified five-brane.   

\acknowledgements

I want to thank David Berman, Feike Hacquebord, Christiaan Hofman, Erik 
Verlinde and Herman Verlinde for providing explanations, advice and 
discussion. Stichting FOM is acknowledged for financial support.


\begin{thebibliography}{88}
\bibitem{ht} C. Hull, P. Townsend, \textit{Unity of Superstring 
Dualities}, \NPB{438}{95}{109}, \hep{9410167}.
\bibitem{vafasen} A. Sen, \textit{A Note on Marginally Stable Bound
States in Type II String Theory}, \PRD{54}{96}{2964},
\texttt{hep-th/9510229}; A. Sen, \textit{U Duality and
Intersecting Dirichlet Branes}, \PRD{53}{96}{2874},  
\hep{9511026}; C. Vafa, \textit{Gas of D-Branes
and Hagedorn Density of BPS States},
\NPB{463}{96}{415}, \hep{9511088}.
\bibitem{polchwitten} J. Polchinski, \textit{Dirichlet Branes and
Ramond-Ramond Charges}, \PRL{75}{95}{4724}, \hep{9510017}; E. Witten,
\textit{Bound States of Strings and P-Branes}, \NPB{460}{96}{335},
\hep{9510135}.
\bibitem{tay} W. Taylor IV, \textit{D-Brane Field
Theory on Compact Spaces}, Phys. Lett. \textbf{B394} (1997) 283,
\texttt{hep-th/9611042}. O.J. Ganor, S. Ramgoolam and W. Taylor IV,
\textit{Branes, Fluxes and Duality in M(atrix) Theory}, Nucl. Phys.
\textbf{B492} (1997) 191, \texttt{hep-th/9611202}. 
\bibitem{piolobers} N.A. Obers, B. Pioline, \textit{U Duality and M
Theory}, \hep{9809039}; S. Elitzur, A. Giveon, D. Kutasov, E. Rabinovici,
\textit{Algebraic Aspects of Matrix Theory on $T^D$}, Nucl. Phys.
\textbf{B509} (1998) 122, \texttt{hep-th/9707217}; B. Pioline, E. 
Kiritsis, \textit{U-duality
and D-brane Combinatorics}, Phys Lett. \textbf{B418} (1998) 61,
\texttt{hep-th/9710078}; N.A. Obers, B. Pioline, E. Rabinovici,
\textit{M-Theory
and U-duality on $T^d$ with Gauge Backgrounds},  
\NPB{525}{98}{163}, \texttt{hep-th/9712084}; M. Blau, M. O'Loughlin, 
\textit{Aspects of U-Duality in
Matrix Theory}, \NPB{525}{98}{182}, \texttt{hep-th/9712047}.
\bibitem{hacq} F.
Hacquebord, H. Verlinde, \textit{Duality Symmetry of ${\cal N}=4$  
Yang-Mills Theory on $T^3$}, Nucl.Phys. \textbf{B508} (1997), 609,
\texttt{hep-th/9707179}. 
\bibitem{bfss} T.
Banks, W. Fischler, S.H. Shenker and L. Susskind, \textit{M-Theory as a
Matrix Model: A Conjecture}, Phys. Rev. \textbf{D55} (1997) 5112,
\texttt{hep-th/9610043}. 
\bibitem{leigh} R. Leigh, \textit{Dirac-Born-Infeld Action from Dirichlet
Sigma Models}, Mod.Phys.Lett. \textbf{A4} (1989) 2767. 
\bibitem{tseytlin} A. Tseytlin, \textit{On
Nonabelian Generalisation of Born-Infeld Action in String Theory},  
\NPB{501}{97}{41}, \hep{9701125}.
\bibitem{brecher} D. Brecher, \textit{BPS
States of the Non-Abelian Born-Infeld Action}, 
\PLB{442}{98}{117}, \hep{9804180}. \bibitem{hvz} C. Hofman, E. Verlinde, G.
Zwart, \textit{U-Duality Invariance of the Four-dimensional Born-Infeld
Theory}, \JHEP{10}{98}{020}, \texttt{hep-th/9808128}. 
\bibitem{aps} M. Aganagic, C. Popescu,
J.  Schwarz, \textit{Gauge-Invariant and Gauge-Fixed D-Brane Actions},
\NPB{495}{97}{99}, \hep{9612080}. 
\bibitem{dvv} R. Dijkgraaf,
E. Verlinde, H. Verlinde, \textit{BPS Quantization of the Five-Brane},
Nucl.Phys. \textbf{B486} (1997) 89, \texttt{hep-th/9604055}.
\bibitem{ow} D. Olive,
E. Witten, \textit{Supersymmetry Algebras that Include Topological
Charges}, \PLB{78}{78}{97} 
\bibitem{matrixstring} R. Dijkgraaf, E. Verlinde, H. Verlinde, 
\textit{Matrix String Theory}, \NPB{500}{97}{43}, \hep{9703030}.
\bibitem{bss} T. Banks, N. Seiberg, S. Shenker, \textit{Branes from 
Matrices}, \NPB{490}{97}{91}, \hep{9612157}.
\bibitem{igusaconforto} J.-I. Igusa, \textit{Theta 
Functions},
Springer, 1972, p 71; Conforto, \textit{Abelsche Funktionen und
Algebraische Geometrie}, Springer, 1956, p76.  
\bibitem{hen} X. Bekaert, M. Henneaux, \textit{Comments on Chiral 
$p$-Forms}, \hep{9806062}.
\bibitem{ST} D. Sorokin, P. Townsend, \textit{M-theory Superalgebra from 
the M-5-brane}, \PLB{412}{97}{265}, \hep{9708003}.
\bibitem{PST} P.Pasti, D. Sorokin, M. Tonin, \textit{Covariant Action for 
a $D=11$ Five-brane with the Chiral Field}, 
\PLB{398}{97}{41}, \hep{9701037}.
\end{thebibliography}
\end{document}